\documentclass{PoS}

\title{Probing parsec scale jets in AGN with geodetic VLBI}

\ShortTitle{Probing parsec scale jets in AGN with geodetic VLBI}

\author{\speaker{Alexander~B.~Pushkarev}%
         \thanks{{\bf Acknowledgements.} This work is based on the analysis of global VLBI observations
         including the VLBA, the raw data for which were provided to us by the NRAO archive. The 
	 National Radio Astronomy Observatory is a facility of the National Science Foundation operated 
	 under cooperative agreement by Associated Universities, Inc. This research has made use of the
	 NASA/IPAC Extragalactic Database (NED) which is operated by the Jet Propulsion Laboratory, 
	 California Institute of Technology, under contract with National Aeronautics and Space 
	 Administration. Y.Y.~Kovalev is a Research Fellow of the Alexander von Humboldt
         Foundation.
}\\
        Max-Planck-Institut f\"ur Radioastronomie, Auf dem H\"ugel 69, 53123 Bonn, Germany \\
	Pulkovo Observatory, Pulkovskoe Chaussee 65/1, St. Petersburg 196140, Russia\\
	Crimean Astrophysical Observatory, Nauchny 98409, Crimea, Ukraine\\
        E-mail: \email{apushkar@mpifr-bonn.mpg.de}}

\author{Yuri~Y.~Kovalev\\
        Max-Planck-Institut f\"ur Radioastronomie, Auf dem H\"ugel 69, 53123 Bonn, Germany\\
	Astro Space Center of Lebedev Physical Institute, Profsoyuznaya 84/32, Moscow 117997, Russia\\
        E-mail: \email{ykovalev@mpifr-bonn.mpg.de}}

\abstract{We report on an ongoing effort to image active galactic nuclei
simultaneously observed at 2.3 and 8.6~GHz in the framework of a
long-term VLBI project RDV (Research \& Development -- VLBA) started
in 1994 aiming to observe compact extragalactic radio sources in the 
astrometric/geodetic mode. Observations of bright extragalactic sources 
are carried out bi-monthly making up to six sessions per year with 
participation of all ten VLBA antennas and up to nine additional 
(geodetic and EVN) radio telescopes. Analysis of single-epoch results 
for 370 quasars, BL Lacs and radio galaxies is presented. We discuss 
VLBI core properties (flux densities, sizes, brightness temperatures), 
spectral characteristics of the cores and jets, evolution of brightness 
temperatures in the jets.
}

\FullConference{The 9th European VLBI Network Symposium on The role of VLBI in the
Golden Age for Radio Astronomy and EVN Users Meeting\\
                 September 23-26, 2008\\
                 Bologna, Italy}

\begin{document}

\section{Introduction}

Long-term VLBI project RDV (Research \& Development -- VLBA) aimed at observations 
of bright compact extragalactic radio sources was started in 1994 under coordination 
of NASA and NRAO \cite{Petrov08}. The simultaneous observations at 2.3 and 8.6~GHz 
are carried out bi-monthly making up to five-six sessions per year with participation 
of all ten 25-m VLBA antennas and up to nine geodetic and EVN stations. The participation 
of the southern antennas such as HartRAO (South Africa) and TIGO (Chile) allowed to 
successfully observe sources with declination up to $-47^\circ$. Sample of observing 
objects consist of $\sim$500 sources. It is important to note that the sample is not 
flux density complete. In each experiment 80-90 active galactic nuclei are observed and 
$\sim$50 objects form a core of the sample and are scheduled continuously. The sample 
is dominated by quasars, with the weak-lined BL Lacs and radio galaxies making up 
8.3\% and 7.8\% of the sample, respectively.

\section{Data Reduction}

The data were correlated at the VLBA correlator in Socorro, with a 4~sec integration 
time, and were obtained by us from the public NRAO archive and then calibrated in 
AIPS using techniques adopted for sub-arrayed data sets. System temperatures and 
values of SEFD measured during the observations were used for the initial amplitude calibration.
Using well measured gains for the VLBA antennas we were able to improve the amplitude calibration 
for non-VLBA stations applying 
self-calibration.
We estimate 
the accuracy of amplitude calibration to be better than 10\%. Phase corrections 
for residual delays and delay rates were done using the AIPS task FRING applying a point-like 
source model. Self-calibration, hybrid mapping, and model fitting were performed in DIFMAP. 
In the model fitting, we used a minimum number of circular Gaussian components that was 
reproducing adequately the observed interferometric visibilities.

\begin{figure}[b]
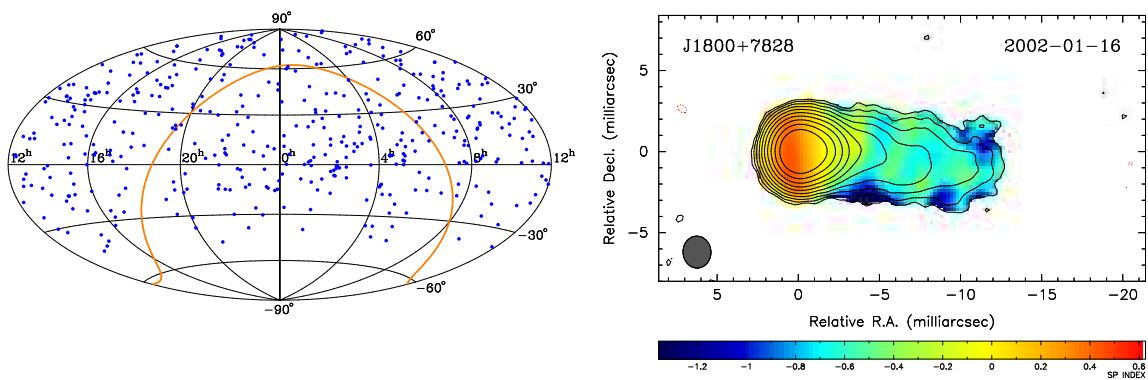

\includegraphics[height=.5\textwidth,angle=-90,clip=true]{figure1a.eps}
\includegraphics[height=.5\textwidth,angle=-90,clip=true]{figure1b.eps}
\caption{Sky distribution for the 370 sources (left). Spectral index distribution in
J1800+7828 calculated between 2.3 and 8.6~GHz with the 8.6~GHz total intensity contours
superimposed (right).}
\label{fig1}
\end{figure}

\section{Results and Discussion}

We discuss first-epoch results for 370 active galactic nuclei (Fig.~\ref{fig1}, left) 
on the basis of high dynamical range ($\sim$1000) images obtained at 2.3 and 8.6~GHz. 
The sources are bright (in 92\% the correlated flux density was more than 200~mJy).
One half of the sources are compact and core-dominated (VLBI compactness greater than 0.51,
core dominance greater than 0.75). The median values of the core size 
are 0.28 and 1.04~mas at 8.6 and 2.3~GHz, respectively. 

Dual-frequency VLBI observations provide a possibility to study spectral properties on parsec 
scales. On Fig.~\ref{fig1} (right) we plot spectral index distribution map for J1800+7828 as 
a typical example. Most of VLBI cores have flat spectra ($\alpha_{\rm core}\sim0$, $S\propto\nu^\alpha$) 
since the radiation from these regions is dominated by optically thin emission from the jet base.
We have cross-identified 48 jet components in 38 sources, the median value of obtained
spectral indices is $\alpha_{\rm jet}=-0.75$ indicating optically thin radiation. This results 
in the median value of power index energy distribution of radiative particles to be $2.5$. 
Determination of spectral index of the jet components has been done taking into account a 
correction for the respective core shifts found in these sources~\cite{Kovalev08}.

\begin{figure}
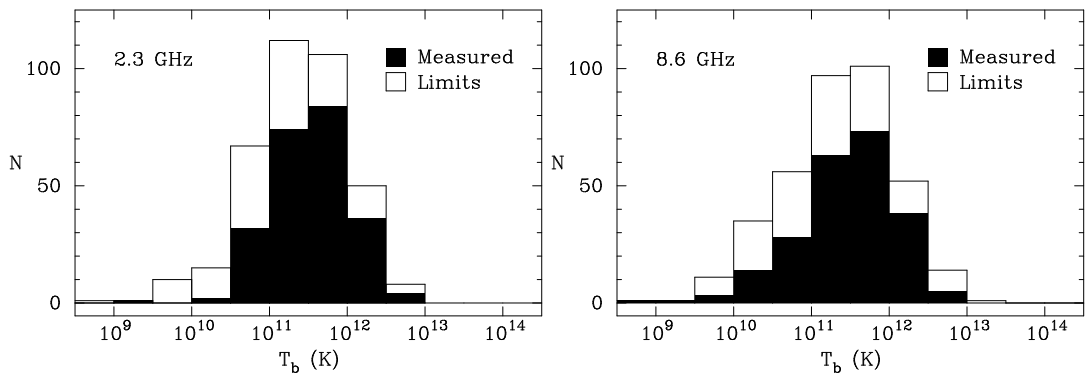

\centering
\includegraphics[height=.47\textwidth,angle=-90]{figure2a.eps}
\includegraphics[height=.47\textwidth,angle=-90]{figure2b.eps}
\caption{Brightness temperature histograms for the core components at 2.3~GHz
(left) and 8.6~GHz (right).}
\label{fig2}
\end{figure}

We have also measured the core brightness temperatures in the source rest frame.
The respective distributions at 2 and 8~GHz shown in Fig.~\ref{fig2} have close median 
values of $2.5\times10^{11}$K. The empty bins are the lower limits and
represent the cases when either the source has unknown redshift or an upper resolution 
limit has been used for the size of the component.

The sources with the prominent and well modeled jets having at least three jet components 
at both frequencies were the cases of our particular interest, since they provided us with 
information about brightness temperature evolution along the jets as a function of a 
distance to the core, $r$. The brightness temperature gradients can be well fitted with power 
law $T_{\rm b}\propto r^{-k}$ for 12 selected sources. The power index $k$ varies with values 
typically between $1.2$ and $3.6$ with the median value of $k=2$. Applying synchrotron 
radiation theory for conical jet model~\cite{Lobanov98,Kadler05} and taking into account 
the median values for jet spectral index $\alpha=-0.75$ and power law index $k=2$ we 
obtained the dependencies of electron density $n_e\propto r^{-1.5}$ and magnetic field 
$B\propto r^{-0.9}$ along the jet.


\begin{thebibliography}{99}

  \bibitem{Petrov08}  L.~Petrov, D.~Gordon, J. Gipson, et al., \emph{Journal of Geodesy}, in press (2009); astro-ph/0806.0167

  \bibitem{Kovalev08} Y.Y.~Kovalev, A.P.~Lobanov, A.B.~Pushkarev, and J.A.~Zensus, 
                      \emph{Opacity in compact extragalactic radio sources and its
		      effect on astrophysical and astrometric studies}, \emph{A\&A}
		      {\bf 483}, 759 (2008).

  \bibitem{Lobanov98} A.P.~Lobanov, \emph{Ultracompact jets in active galactic nuclei}, 
                      \emph{A\&A} {\bf 330}, 79 (1998).

  \bibitem{Kadler05}  M.~Kadler, \emph{PhD thesis}, Bonn University (2005).
\end{thebibliography}
\end{document}